\documentclass{iopjournal}

\usepackage{lineno}
\usepackage[utf8]{inputenc}
\usepackage{graphicx}
\usepackage{pdfpages}
\usepackage{epstopdf}
\usepackage{subfig}
\usepackage{amsmath,amssymb}

\begin{document}
\articletype{Paper} 

\title{\boldmath Effects of Cosmic Muons on $\mu$eV-to-meV Scale Axion Dark Matter Searches}

\author{Dan Zhang$^{1,*}$\orcid{0000-0002-2910-5805}, Gray Rybka$^1$\orcid{0000-0001-9973-1564}, Edward J. Daw$^2$\orcid{0000-0002-3780-5430} and Robyn Evren$^2$\orcid{0009-0001-7370-1399}}

\affil{$^1$University of Washington, Seattle, Washington 98195, USA}

\affil{$^2$University of Sheffield, Sheffield S10 2TN, UK}

\affil{$^*$Author to whom any correspondence should be addressed.}

\email{dzhang95@uw.edu,grybka@uw.edu,e.daw@sheffield.ac.uk,revren1@sheffield.ac.uk}


\begin{abstract}
{We estimate the synchrotron radiation of cosmic muons in a uniform magnetic field in the $\mu$eV-to-meV energy scale. Such events can potentially bring backgrounds to the axion dark matter searches. The GEANT4 software package is utilized to simulate the muon tracks in a cylindrical region of interest with an 8~T solenoid magnetic field applied. We further develop an analytical estimation of the angular-frequency-differential synchrotron radiation power spectra in this work as the cosmic muons span a wide range of Lorentz factor $\gamma$ and pitch angle $\alpha$. We verify that the cosmic muons are not the dominant noise background for the current axion dark matter experiments on the $\mu$eV scale because of the high quality factor $Q$ and fine energy resolution in the readout. However, without sufficient energy resolution in the detector readout, future broadband axion dark matter experiments will be vulnerable to the synchrotron radiation of these charged particles.}

\end{abstract}

\section{Introduction}
\label{sec:intro}
The Quantum chromodynamics (QCD) axion is a compelling dark matter candidate that can explain 100\% of the dark matter abundance occupying 85\% mass of the universe and simultaneously solve the strong CP problem at the same time~\cite{Peccei1977June,Weinberg:1977ma,Wilczek:1977pj,PRESKILL1983127,ABBOTT1983133,DINE1983137,PhysRevLett.50.925}.  The minimal model is that the QCD axions are the pseudo–Nambu–Goldstone boson of the $U(1)_{\text{PQ}}$ proposed by Roberto Peccei and Helen Quinn to explain the CP conservativeness within the strong interaction~\cite{Peccei1977June}. The $U(1)_{\text{PQ}}$ stays as one of the most promising solutions to the strong CP problem. As an extension of the strong interaction, the QCD axion can be naturally produced in the early universe when the temperature drops to the energy scale of the $U(1)_{\text{PQ}}$ spontaneous symmetry breaking $f_a$~\cite{Borsanyi2016,PhysRevD.96.095001}.

The electroweak scale $f_a$ was quickly ruled out after being proposed~\cite{BARDEEN1978229,PhysRevD.18.1607}. Stellar evolution puts a lower limit on $f_a\gtrsim10^8$~GeV~\cite{Dolan_2022,PhysRevLett.113.191302}, the upper limit on the QCD axion mass, $m_a\lesssim1$~eV. For QCD axion dark matter, a stronger upper limit $m_a\lesssim0.1$~eV recently has been reported using data from the James Webb Telescope~\cite{pinetti2025constraintsqcdaxiondark}. Post-inflation axion dark matter production mechanisms, such as the misalignment mechanism and topological defect decays, strongly motivate the $\mu$eV-to-meV energy scale~\cite{PhysRevD.83.123531,PhysRevD.91.065014,PhysRevD.92.034507,Fleury_2016,Bonati2016,PETRECZKY2016498,Borsanyi2016,PhysRevLett.118.071802,Klaer_2017,PhysRevD.96.095001,PhysRevLett.124.161103,10.21468/SciPostPhys.10.2.050,buschmann2021dark}. Pre-inflation production with the misalignment mechanism broadens the motivated lower mass border to 0.1~$\mu$eV~\cite{PRESKILL1983127,ABBOTT1983133,DINE1983137}, and the stochastic mechanism provides possibilities for even lighter axion dark matter masses~\cite{PhysRevD.98.035017,PhysRevD.98.015042}.

At such a low-energy scale, only Sikivie Haloscopes \cite{sikivie1983} have reached to the benchmark models of QCD axion dark matter: the Kim-Shifman-Vainshtein-Zakharov (KSVZ)  axion~\cite{Kim:1979if,Shifman:1979if} and the Dine-Fischler-Srednicki-Zhitnitsky (DFSZ) axion~\cite{Dine:1981rt,Zhitnitsky:1980tq}. A Sikivie Haloscope measures axion-to-two-photon coupling $g_{a\gamma\gamma}$ with a resonant cavity in a high magnetic $B$ field. The $B$ field supplies virtual photons to stimulate the conversion of axion dark matter to photons. The photons emitted from the conversion, carrying the total energy of the axion $m_a$, is confined in the resonant cavity and extracted out by an antenna or lost to the wall. 

Currently, the most sensitive axion haloscopes lie in the frequency range 0.5 to 1.5~GHz (2 to 6~$\mu$eV)~\cite{admx2018,admx2020,admx2021,admx2024,admx2025,capp2023,capp2024}. In this radio frequency to microwave range, the dominant noise that limits the axion sensitivity is typically Johnson noise of the resonator and the electronic noise of the readout~\cite{PhysRevD.111.092012}. Natural radioactivities are not traditionally considered as backgrounds.

However, the radiation of cosmic muons can induce some physical events that may become noise to axion dark matter searches in a haloscope. 
Charged particles moving through the magnetic $B$ field can generate cyclotron or synchrotron radiations confined in the resonator, contaminating the axion dark matter search. Since the radiation power is proportional to $B^2$, the same as the axion signal power, this noise power can be crucial if large enough. To the authors' knowledge, there is no publication on estimating the effects of natural radioactivities and cosmic rays on the axion dark matter haloscopes.

In this manuscript, we explore the synchrotron radiation of cosmic muons at sea level passing through a 8~T uniform solenoid magnetic $B$ field, representing a typical axion haloscope magnetic field. We first use the GEANT4 package~\cite{geant4_1,geant4_2,geant4_3} to simulate the trajectories of cosmic muons. Due to the non-vertical pitch angle of the cosmic muons to the $B$ field and the non-negligible ratio of the muons with moderate Lorentz factors, $\gamma<10$, the GEANT4 synchrotron radiation package is inappropriate for our simulation. Therefore, we integrate the total synchrotron radiation analytically~\cite[pg. 177-181]{1966}. 

The simulation results show that such backgrounds are negligible for the current axion haloscope experiments including ADMX and CAPP because of the signal enhancement from the high quality factor $Q$ of the copper cavity resonators and the fine energy resolution in the readout. However, for future broadband axion experiments, such as BREAD and MADMAX~\cite{bread,madmax}, if the readout contains photon counters without high enough energy resolution, the synchrotron radiation of cosmic muons might limit the sensitivity.

\section{Method}

The full simulation can be divided into three steps: cosmic muon generation, trajectory recording and analytical synchrotron radiation integration. 

The cosmic muon generator developed in Ref.~\cite{cosmicMuonSource} is integrated into the GEANT4 framework as the particle source. It provides both flat-plane and cylindrical side-wall generations, which simplifies our simulation of cosmic muons entering the bore of a solenoid magnet. We define the region-of-interest to be a copper cavity with 1~m length and 0.2~m radius inside a solenoid magnet bore, which is similar to the facility at ADMX~\cite{admxDetails}. For such a geometry, we expect a total flux of $17$~muons/s going through the top flat surface, and $40$~muons/s passing from the side-wall at sea level~\cite{cosmicMuonSource}.
We consider these two individual groups of muons separately.

The cosmic muons are generated in the simulation slightly outside this region of interest, being 1~mm away from the outside wall of the copper cavity. The low-energy muons with small bending radius cannot enter the cylindrical cavity and are naturally filtered out. The timestamps of muons entering and leaving the cavity, together with their pitch angles $\alpha$ and Lorentz factors $\gamma$ are the most important information obtained from the GEANT4 simulation. 
Figure 1 visualizes 10 trajectories of cosmic muons using the GEANT4 simulation package.

 \begin{figure}[htbp]
    \centering
    \subfloat[10 trajectories entering from the top]{\includegraphics[width=0.5\linewidth]{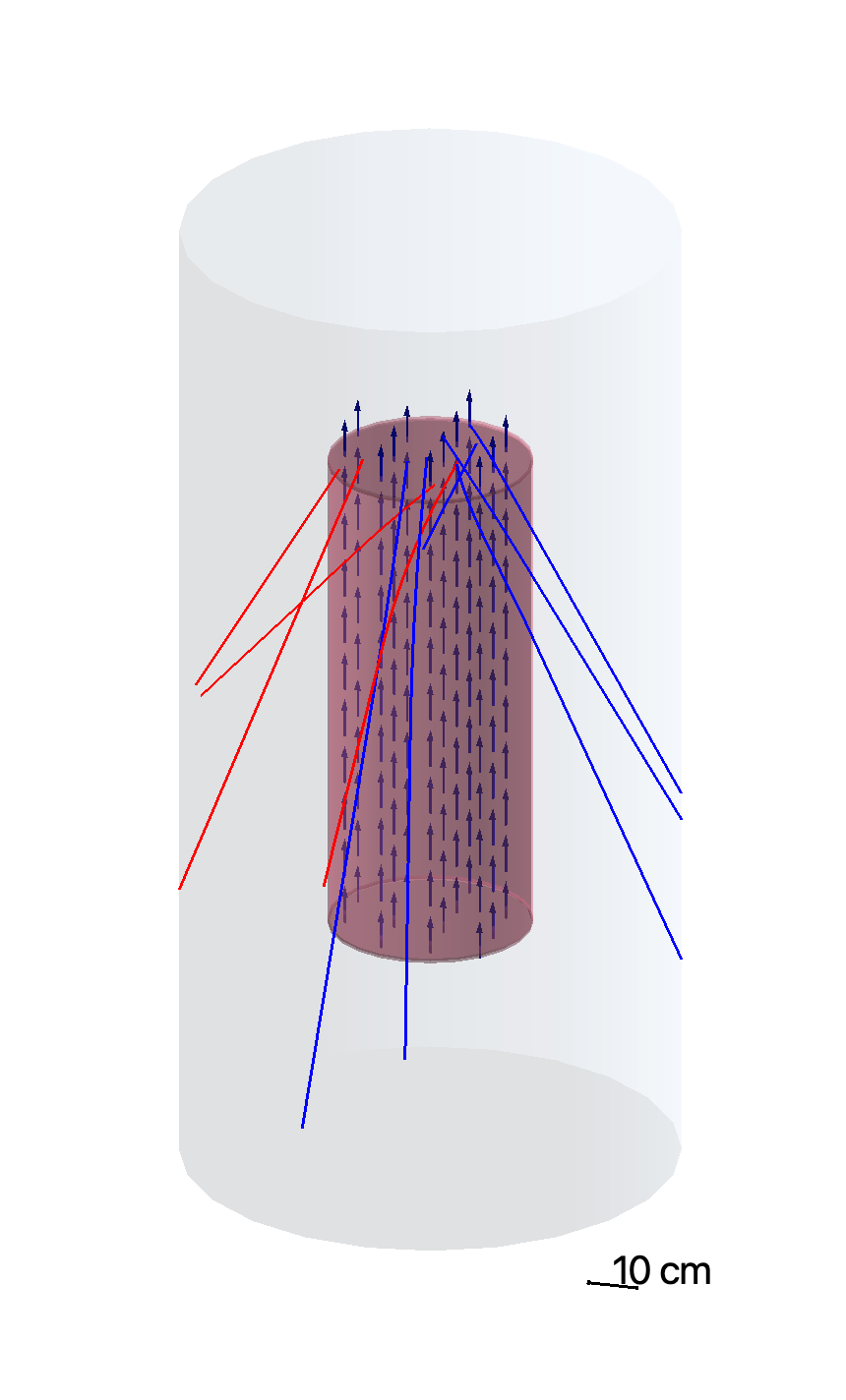}}
    \hfill
    \subfloat[10 trajectories entering from the wall]{\includegraphics[width=0.5\linewidth]{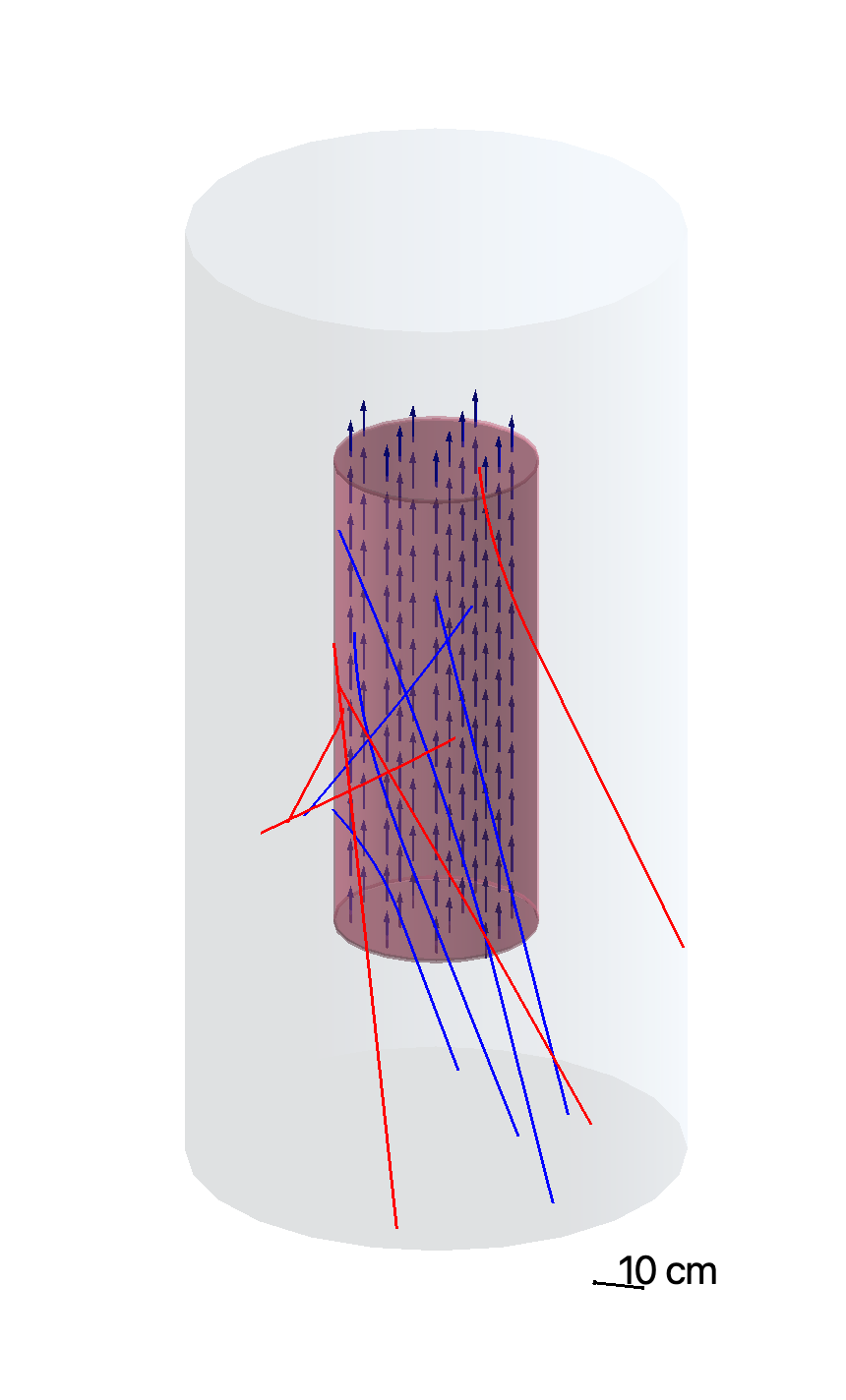}}
    \caption{Examples of cosmic muon trajectories passing through a copper cavity with GEANT4. Blue and red tracks are $\mu^{+}$ and $\mu^{-}$, respectively. An 8~T vertical magnetic field (dark blue arrows) along $z$-axis is applied to cover the whole copper cylindrical cavity (pink). A larger cylindrical space (grey) is defined to terminate the events.}
    \label{fig:geant4simu}
\end{figure}

In this work, we estimate the angular-frequency-differential  synchrotron radiation power spectra in the cylinder from $\mu$eV to meV.
We develop our own synchrotron radiation integration  based on Ref.~\cite{1966}.
The energy spectrum of cosmic muons at sea level is broad, where 23\% are $\gamma<10$ and 13\% are $\gamma>100$ (Fig.~\ref{fig:alpha_gamma_pdf}). The pitch angles of the muons entering the magnetic field also spread all possibilities from 0$^{\circ}$ to 90$^{\circ}$. A detailed understanding of how the synchrotron radiation changes with different pitch angles is necessary for our calculation.

Here we present more details of the non-trivial synchrotron radiation theories. The total  radiation of a charged particle can be calculated following Liénard's formula~\cite[pg. 664-668]{lienard}:
\begin{align}\label{equ:lienard}
P &= \frac{q^2}{6\pi\epsilon_0 c^3} \gamma^6 \left( \dot{\vec{\beta}}^2 - (\vec{\beta}\times \dot{\vec{\beta}})^2 \right) \nonumber\\
& = \frac{q^4 B^2 \beta^2 \gamma^2 \sin^2 \alpha}{6 \pi \epsilon_0 c m^2},
\end{align}
where $q$ and $m$ is the charge and mass of the particle, respectively, $\vec{\beta}=\vec{v}/c$ is the normalized velocity and $\alpha$ is the pitch angle between $\vec{\beta}$ and $\vec{B}$. SI units are used where $\epsilon_0$ is the vacuum permittivity and $c$ is the speed of light.

\begin{figure}
    \centering
    \includegraphics[width=0.5\linewidth]{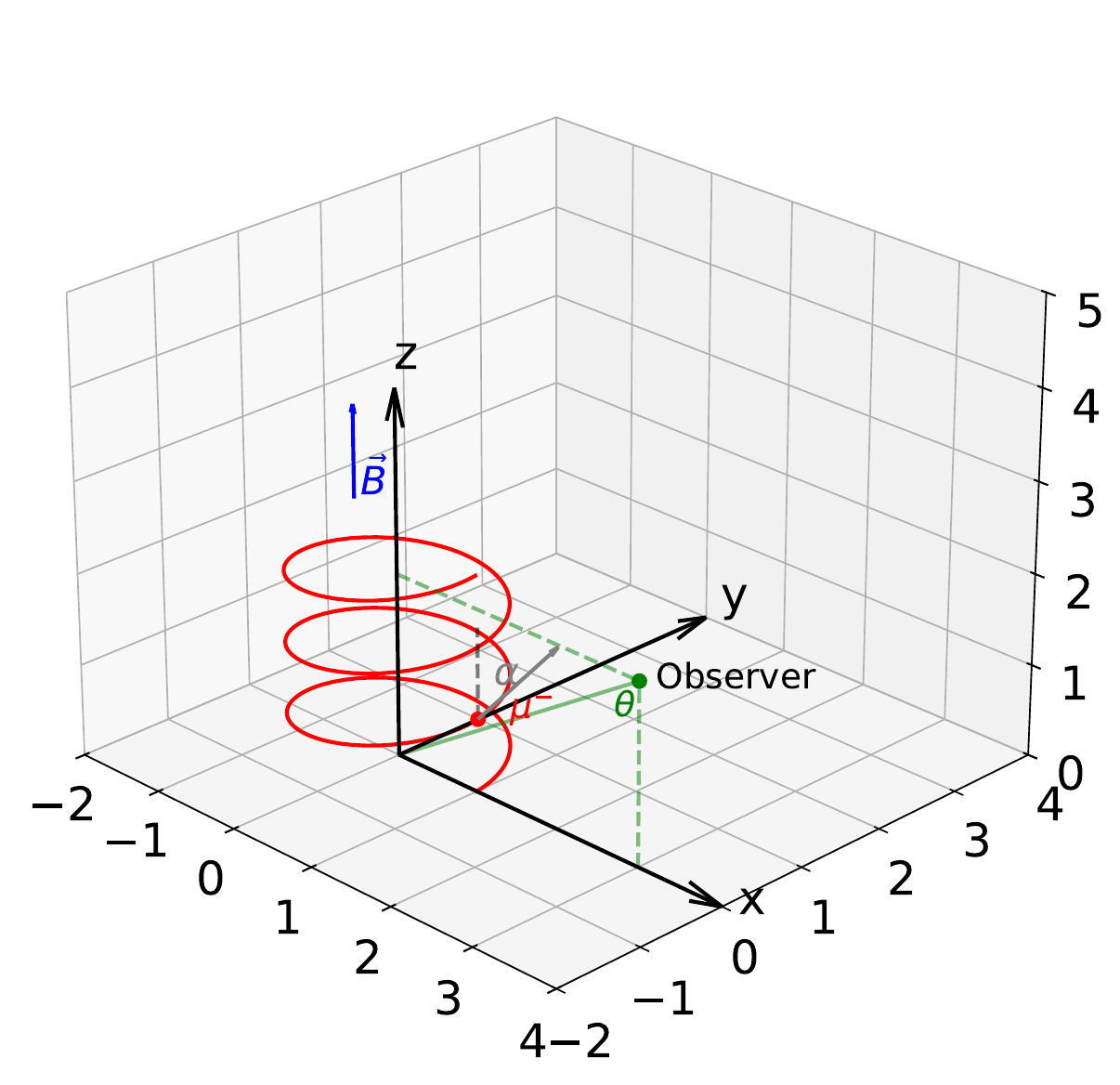}
    \caption{Vector diagram for a muon in helical motion in a uniform magnetic field. Both $\alpha$ and $\theta$ are defined in the lab frame.}
    \label{fig:coordinate}
\end{figure}

The total radiation $P$ in Eq.~\ref{equ:lienard} is used as a sanity check of our numerical integration based on Ref.~\cite{1966}:
\begin{equation}\label{equ:totalSynNum}
    P =\frac{q^2 \omega_0^2}{8\pi^2\epsilon_0 c} \sum_{n=1}^{\infty}\Big(\int   p_n(\alpha,\theta)d\Omega\Big).
\end{equation}
Here $n$ is harmonic number of the synchrotron radiation, $\omega_0=qB/m$ is the gyro-frequency and $p_n(\alpha,\theta)$ is:
\begin{equation}\label{equ:emis}
    p_n(\alpha,\theta) = \frac{n^2\beta_{\perp}^2(1-\beta^2)}{(1-\beta_{\parallel}\cos{\theta})^3}\Big([J_n^{'}(n\kappa)]^2 + \big(\frac{\cos\theta-\beta_{\parallel}}{\beta_{\perp}\sin\theta}\big)^2[J_n(n\kappa)]^2\Big),
\end{equation}
where
\begin{align*}
\beta_{\perp}&=\beta\sin\alpha,~~\beta_{\parallel}=\beta\cos\alpha,\\
    \kappa &= \frac{\beta_{\perp}\sin\theta}{1-\beta_{\parallel}\cos{\theta}}.
\end{align*}
$J_n$ and $J^{'}_n$ are the nth order of the Bessel functions and its first derivative, respectively.
Note that Eq.~\ref{equ:emis} already integrates the delta function of the allowed frequencies in the synchrotron radiation. The coordinate selections of the lab frame $\alpha$ and $\theta$ are shown in figure \ref{fig:coordinate}.

The azimuthal symmetry in Eq.~\ref{equ:emis} reduces the solid angle integral$\int d\Omega$ to a one-dimensional integral over the polar angle $\theta$: $\int d\Omega = \int_0^{\pi} 2\pi\sin\theta d\theta$. 
Our numerical integral following Eq.~\ref{equ:totalSynNum} agrees with Eq.~\ref{equ:lienard}
at different pitch angles $\alpha$ and Lorentz factors $\gamma$.    

\begin{figure}
    \centering
    \includegraphics[width=0.5\linewidth]{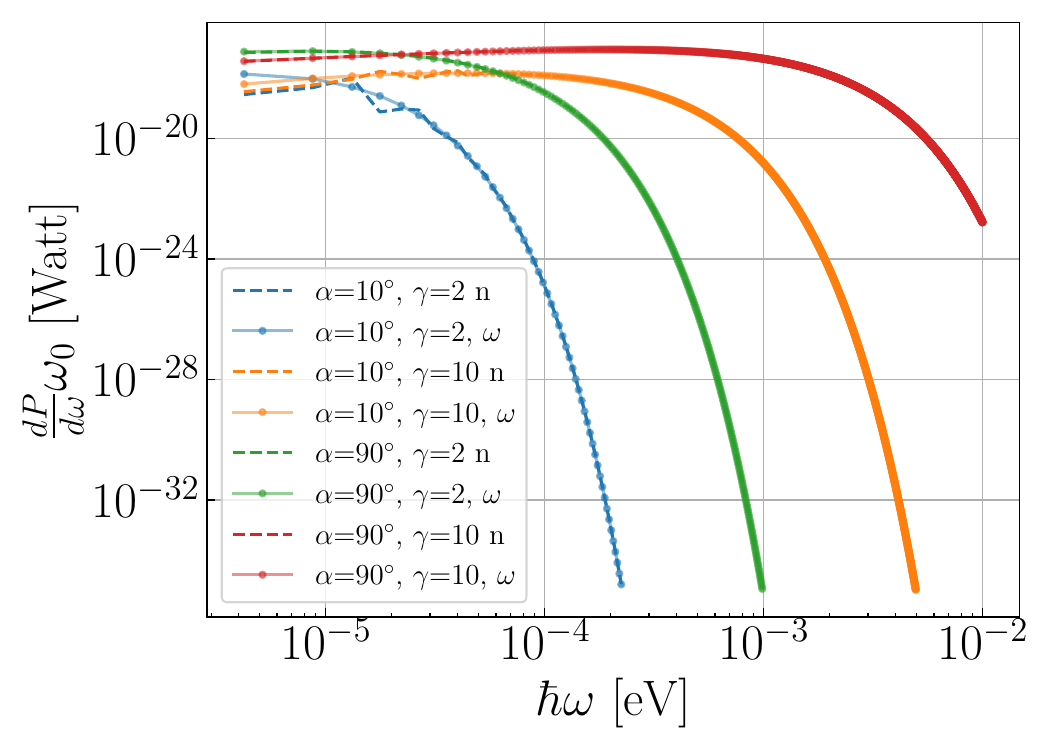}
    \caption{Integrated angular-frequency-differential synchrotron radiation power spectra at different pitch angles $\alpha$ and Lorentz factors $\gamma$. The summations with $n$  (dashed lines) agree with substituted $\omega$ integrals (dotted lines) when the integration window from $E = \hbar\omega$ to $\hbar(\omega+\omega_0)$ spreads a large number of $n$.}
    \label{fig:individualdPdomega}
\end{figure}

To calculate the angular-frequency-differential synchrotron radiation power spectrum $dP/d\omega$, we  replace the harmonic number $n$ with the angular frequency $\omega$ in the lab frame:
\begin{equation}\label{equ:nVSomega}
   n = \gamma \omega  (1-\beta_{\parallel}\cos\theta)/\omega_0.
\end{equation}
When $n\gg1$, $dn \approx \gamma(1-\beta_{\parallel\cos\theta})d\omega/\omega_0 $ and the direct substitution is appropriate in Eq.~\ref{equ:emis}. However, when $n$ is small, the replacement can cause more than a factor of 2 deviation since the continuous integral with $\omega$ is an average calculation while $n$ can only be integers. The deviation at small $n$ for small $\gamma$ can be seen in figure \ref{fig:individualdPdomega}. 

\begin{figure}[htbp]
    \centering
    \subfloat[$E=5\times10^{-6}$~eV, $\omega/2\pi = 1.2$~GHz]{\includegraphics[width=0.5\textwidth]{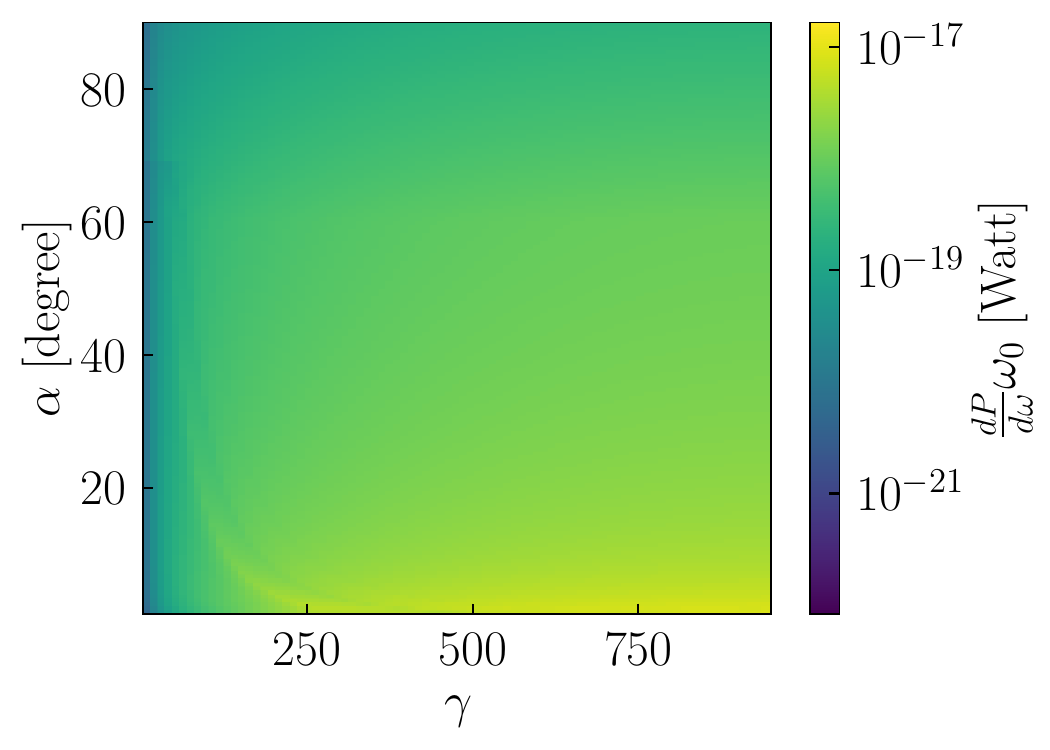}}
    \hfill
    \subfloat[$E=5\times10^{-4}$~eV, $\omega/2\pi = 120.9$~GHz]{\includegraphics[width=0.5\textwidth]{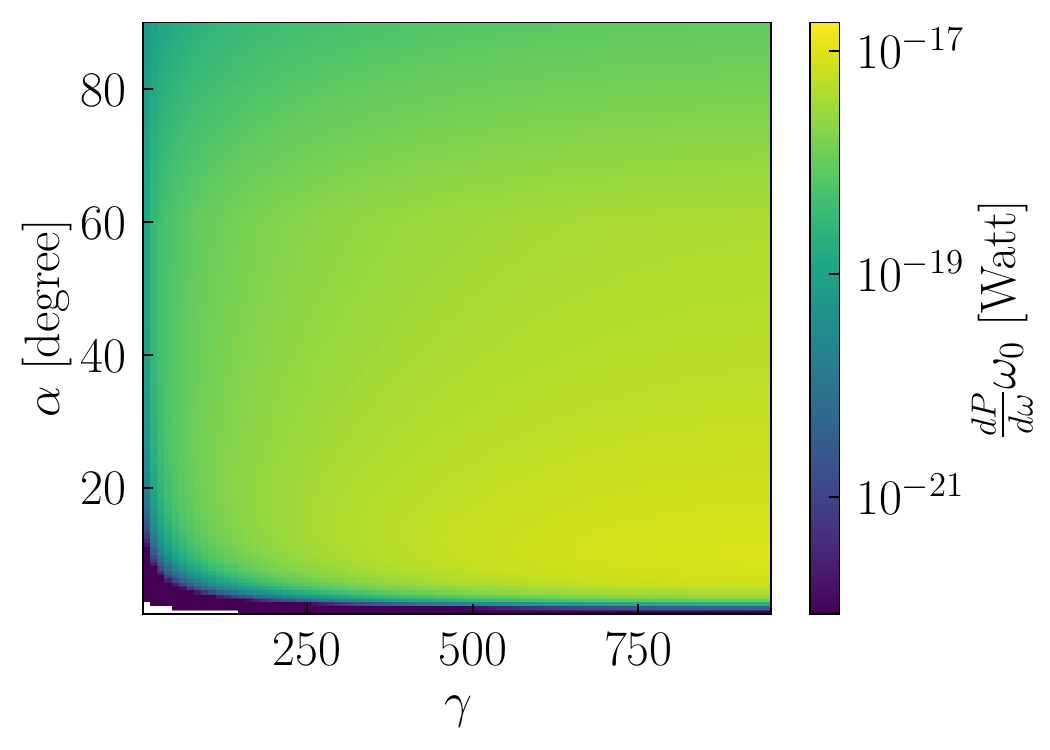}}
    \caption{The angular-frequency-differential synchrotron radiation power spectra at different $\omega$s with the same integral window $\omega_{0}$ as a function of pitch angle $\alpha$ and Lorentz factor $\gamma$.}
    \label{fig:sync_rad_maps}
\end{figure}

Therefore, we implement a hybrid calculation for the angular-frequency-differential synchrotron radiation power spectrum $dP/d\omega$. The angular frequencies are binned for the calculation of $dP/d\omega$. For each bin with a lower bound frequency $\omega_l$ and an upper bound $\omega_u$. The related boundaries on $n$ are $n_l=\gamma\omega_l(1-\beta_{\parallel})/\omega_0$ and $n_u = \gamma\omega_u(1+\beta_{\parallel})/\omega_0
$. When $n_u - n_l > 100$, we replace $n$ with Eq.~\ref{equ:nVSomega} directly for the numerical integral.
Otherwise, we loop over each $n$ and find the corresponding limits on $\theta$ for the integral.
Figure \ref{fig:sync_rad_maps} illustrates some examples of the angular-frequency-differential synchrotron radiation power spectra with different $\alpha$ and $\gamma$. These analytical integrated synchrotron radiation power spectra provide the basic elements of the last step of the simulation. At a specific angular frequency,  the averaged synchrotron radiation is normalized based on the cosmic muon events with different $\gamma$s and $\alpha$s according to the GEANT4 simulation results.

\section{Results}
\begin{figure}[htbp]
    \centering
    \subfloat[Top]{\includegraphics[width=0.5\textwidth]{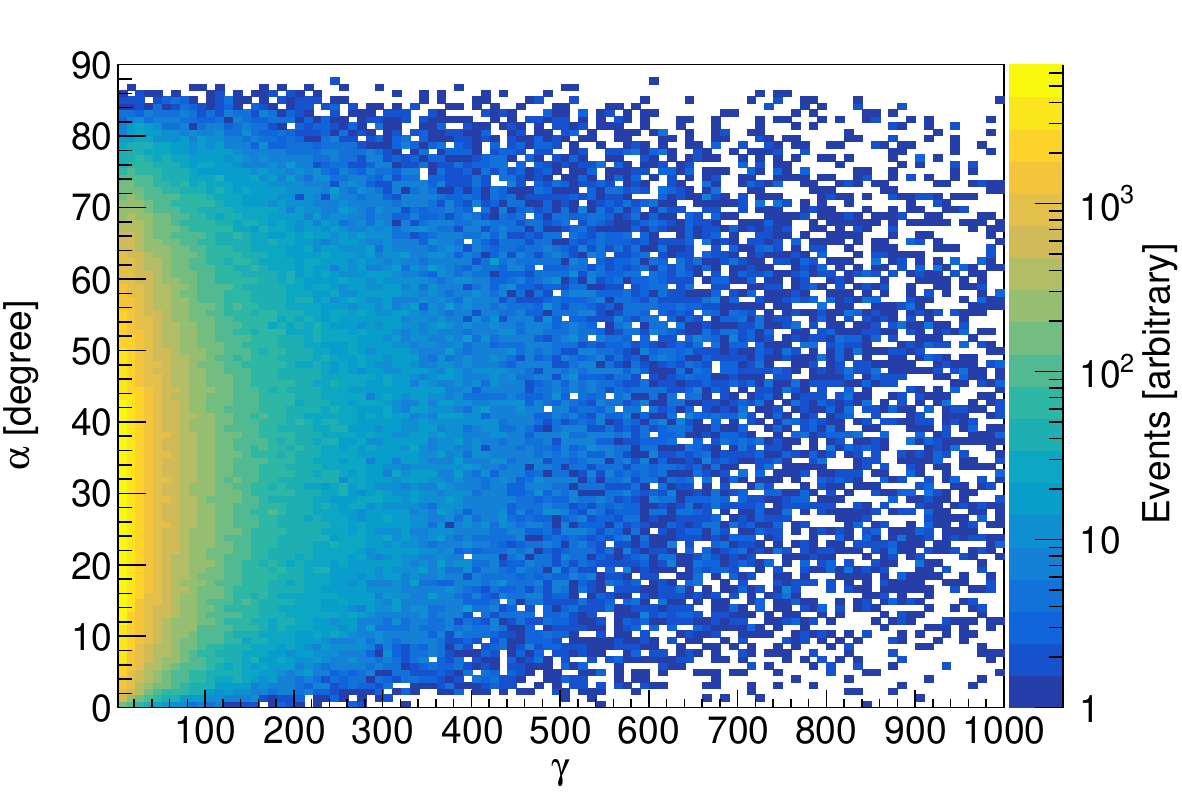}}
    \hfill
    \subfloat[Wall]{\includegraphics[width=0.5\textwidth]{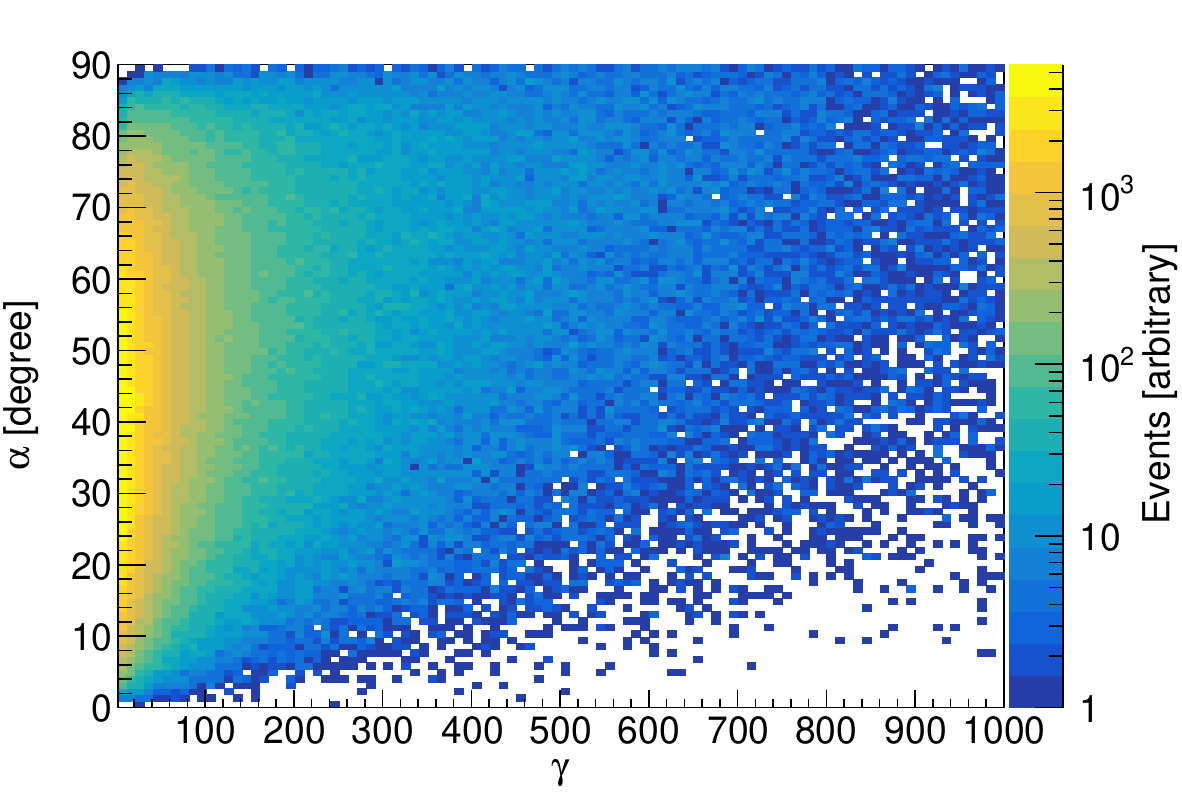}}
    \caption{Pitch angle $\alpha$ and Lorentz factor distributions $\gamma$ of cosmic muons at sea level}.
    \label{fig:alpha_gamma_pdf}
\end{figure}

First, we investigate the pitch angle and Lorentz factor distribution of the cosmic muons entering the cylindrical cavity. Figure \ref{fig:alpha_gamma_pdf} shows the distributions of the muons entering from the top of the cavity in (a) and the wall of the cavity in (b). Due to more muons entering the cavity from the wall, the synchrotron radiation power of the wall muon group is more significant. Note, a lower boundary of $\gamma>2$ was implemented for the simulated muon generator, corresponding to the momentum $p>0.1~$GeV. This is the valid minimum momentum of the cosmic muon generator that agrees with the experimental data in Ref.~\cite{cosmicMuonSource}.

\begin{figure}[htbp]
    \centering
    \subfloat[Top]{\includegraphics[width=0.5\textwidth]{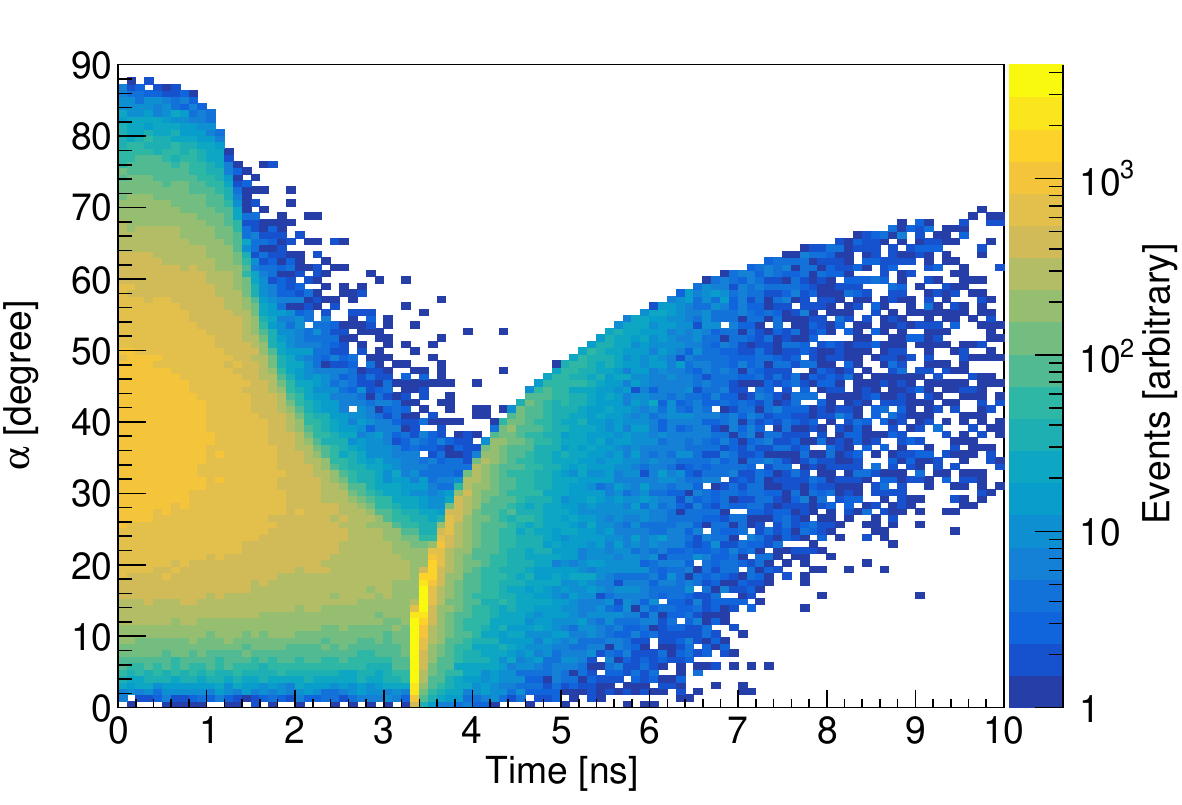}}
    \hfill
    \subfloat[Wall]{\includegraphics[width=0.5\textwidth]{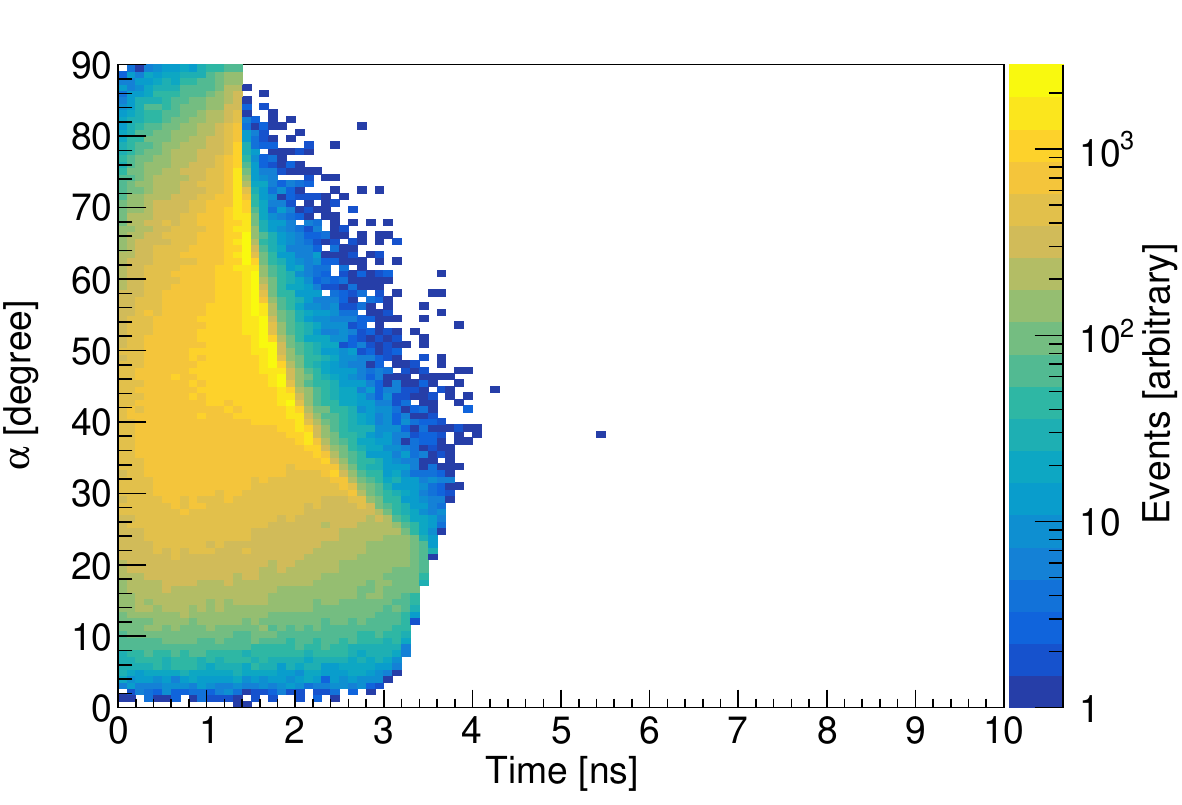}}
    \caption{Distribution of the muon time in the copper cavity with different pitch angles $\alpha$}.
    \label{fig:time_alpha_pdf}
\end{figure}

Figure \ref{fig:time_alpha_pdf} illustrates the time muons stay in the region of interest for both the top and wall groups. Most of the muons stay in the 1~m long cylindrical space for several nanoseconds, which agrees with the speed of light estimation. Interestingly, there is a tail feature in the $\alpha$-time distribution. Further investigation shows that these events are more mildly relativistic with $\gamma<10$, and the cyclotron radius is smaller than the radius of the cavity (0.2~m). They stay longer in the cavity with helical motions as exemplified in figure \ref{fig:coordinate}.

We focus on the synchrotron radiation from the cosmic muons between $4~\mu$eV and 10~meV. This energy range covers most of the well-motivated energy range for  post-inflation QCD axion dark matter.
The gyro-frequency of a muon in 8~T is $\omega_0=6.8\times10^9$~rad/s, which is taken as the start of the angular-frequency-differential synchrotron radiation power spectrum as well as the step in the spectrum. Correspondingly, the energy step size is $\delta E=\hbar\omega_0=4.4~\mu$eV. 

The angular-frequency-differential synchrotron radiation power spectra for different $\alpha$ and $\gamma$ are calculated separately. In the frame where the muon undergoes circular motion,  synchrotron radiation can only be emitted at specific frequencies, $n\omega_0$, corresponding to different harmonic numbers $n$. To visualize the differential synchrotron radiation spectra, we use $\frac{dP}{d\omega}\omega_0$ to present the averaged differential synchrotron radiation over each $\omega_0$ integral window. Figure \ref{fig:sync_rad_maps} presents two examples of the differential synchrotron radiation spectra at 5~$\mu$eV and 0.5~meV as a function of the pitch angle $\alpha$ and Lorentz factor $\gamma$.

\begin{figure}
    \centering
    \includegraphics[width=0.5\linewidth]{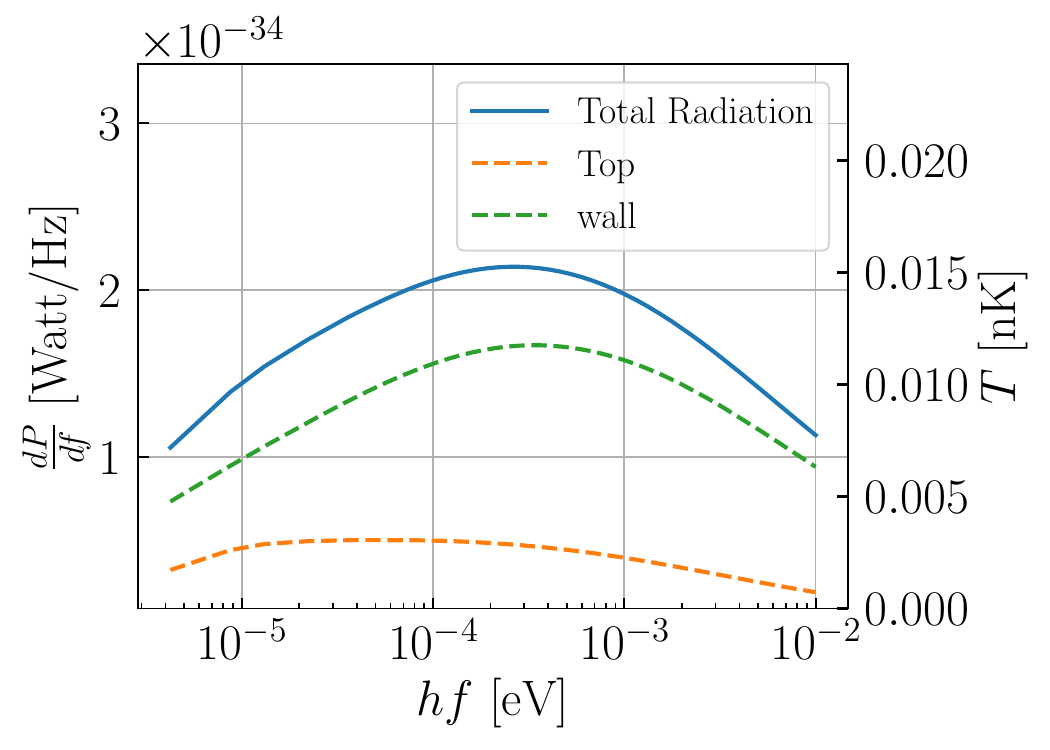}
    \caption{The averaged differential synchrotron radiation power spectrum (blue solid line) of the cosmic muons. The contributions of the top muons (orange dashed line) and the wall (green dashed line) are scaled with the expected rates at the sea level. The $y$-axis on the right-hand side converts the energy into noise temperature $T = \frac{dP}{df}\cdot\frac{1}{k_B}$, where $k_B$ is the Boltzmann constant.}
    \label{fig:averaged_syn_spec}
\end{figure}

The averaged synchrotron radiation power spectra of the cosmic muons over different $\alpha$ and $\gamma$ at sea level are shown in figure \ref{fig:averaged_syn_spec}. We converted the angular frequency $\omega$ to frequency $f = \omega/2\pi$ since the latter is more commonly used in the axion haloscope community.  We have scaled the spectra with the expected muon rates: 17~muons/s for the top and 40~muons/s for the wall. The difference in the muon rates dominates the contribution of the synchrotron radiation. Both the top and the wall present broad radiation, spreading from $\mu$eV to meV with the synchrotron radiation power on the same order.

To explain how the muon synchrotron radiation can affect axion dark matter detection, we compare the muon synchrotron noise power to the microwave signal power from the axion-to-photon conversion of a Sikivie haloscope~\cite{sikivie1983,admxDetails}. The microwave signal power generated inside a haloscope is:
\begin{equation}\label{eqn:axion}
P_{\mathrm{a}}=1.0\times 10^{-27}~\text{W} \left(\frac{V_\text{eff}}{50~\ell}\right)\left(\frac{B}{8~\text{T}}\right)^2 \left(\frac{g_\gamma}{0.36}\right)^2
\left(\frac{\rho_a}{0.45~\text{GeV/cc}}\right)\left(\frac{f}{1~\text{GHz}}\right){Q}.
\end{equation}
Here $V_\text{eff}$ is the effective volume that includes the efficiency according to the overlapping of the external $\vec{B}$ field and the cavity electric field mode at the frequency $f$. The $f$ is the microwave photon frequency carrying the total energy of the axion. For cold dark matter, $f\approx m_ac^2/h$, where $m_a$ is the axion mass. The $g_{\gamma}$ is the model-dependent constant: the KSVZ  axion has $g_{\gamma}=-0.97$~\cite{Kim:1979if,Shifman:1979if} and the more feebly coupled DFSZ axion has $g_{\gamma}=0.36$~\cite{Dine:1981rt,Zhitnitsky:1980tq}. The $\rho_a$ is the local dark matter density~\cite{darkMatterDensity_1,darkMatterDensity_2}.
The qualify factor $Q$ can provide significant enhancement for a Sikivie haloscope. For ADMX at 1~GHz, if $f$ is on resonance with the copper cavity, typically $Q$ can reach $50,000$. 

\begin{figure}[htbp]
    \centering
    \subfloat[Power Spectra]
    {\label{fig:axion_vs_sync_rad_a}
    \includegraphics[width=0.48\linewidth]{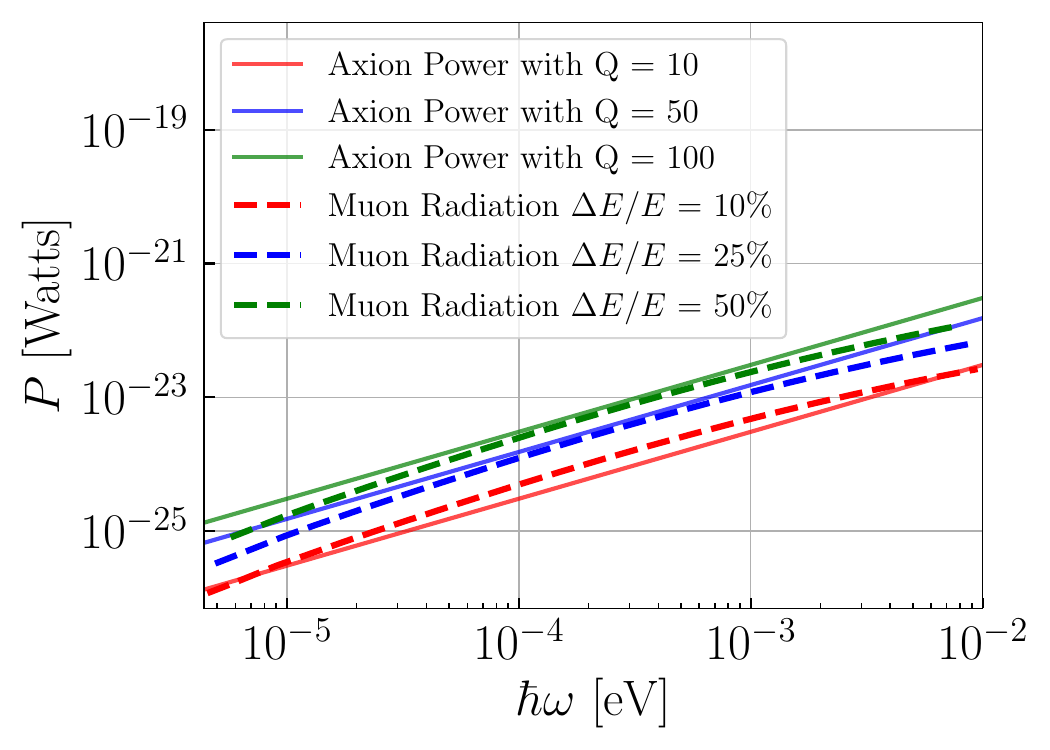}}
    \hfill
    \subfloat[Axion VS Muon Power]
    {\label{fig:axion_vs_sync_rad_b}
    \includegraphics[width=0.48\linewidth]{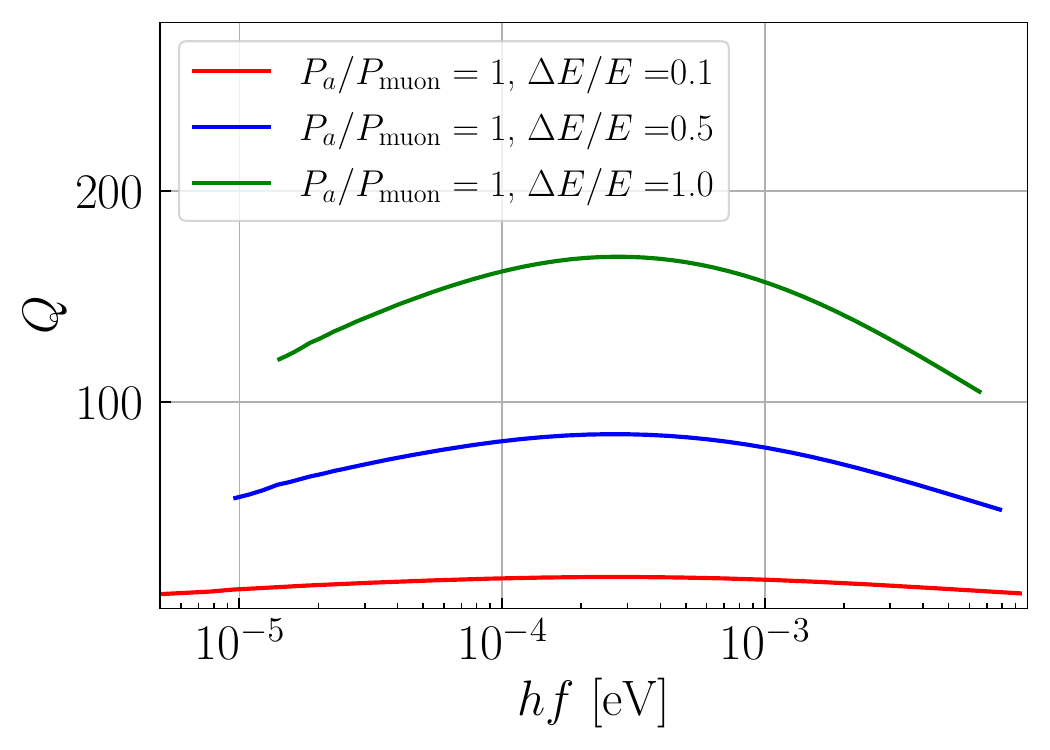}}
    \caption{(a) The axion power $P_a$ with different enhancements $Q$ (solid lines) compared to muon synchrotron radiation with different energy resolutions (dashed lines). Here typical parameters close to the ADMX experiments are used in the $P_a$ calculations~\cite{admxDetails}. (b) The $Q$s leading to $P_{a} = P_\text{axion}$ at different energy resolutions.}
    \label{fig:axion_vs_sync_rad}
\end{figure}

In figure \ref{fig:axion_vs_sync_rad_a}, DFSZ axion power is calculated with  $V_\text{eff}=50~\ell$ and different $Q$s. The muon synchrotron radiation  power is calculated based on the total radiation in figure \ref{fig:averaged_syn_spec} and different assumed relative energy resolutions $\Delta E/E$. 
When $\Delta E/E=10$\%, the muon contribution is comparable to the DFSZ axion signal power at $Q=10$. Figure \ref{fig:axion_vs_sync_rad_b} further illustrates the $Q$  for different energy resolutions if $P_a=P_\text{axion}$, which indicates the minimum $Q$ requirements.  For worse energy resolution, the effective frequency span in our calculation is smaller. The noise power of the muons' synchrotron radiation is trivial for experiments such as ADMX on $\mathcal{O}(1)~\mu$eV. However, when low energy resolution photon detectors are used and lower $Q$ is designed for broader detection band, the cosmic muon synchrotron radiation is likely to be a source of non-negligible noise power.

\section{Discusion and Conclusion}
Cosmic muons are only one type of natural event that brings broadband microwave radiation. Secondary charged particles excited by the muons and beta decays might have similar effects for axion haloscopes. The relativistic electrons can increase the requirements on the energy resolution and the signal enhancement by several orders for broadband experiments, which we plan to investigate in future works.

We only compare the initial synchrotron radiation and signal power inside the cavity. Some detector designs, including BREAD and MADMAX~\cite{bread,madmax}, will only collect the synchrotron radiation with some specific polarizations. The methodology can be adapted to different experiments according to Eq.~\ref{equ:totalSynNum} and \ref{equ:emis} for estimating the synchrotron radiation effects of charged particles.

The design of a Sikivie haloscope can be used for high-frequency gravitation wave searches directly~\cite{gavnet}. Similar designs can also be found in the beta decay measurements using cyclotron radiation emission spectroscopy~\cite{project8,helium6}. The  considerations discussed in this work should therefore be taken into account in the planning and design of these future experiments as well.

In brief, we study the possible noise power coming from cosmic muons for axion dark matter searches between 4~$\mu$eV and 10~meV. 
We explain that cosmic muons and other natural radioactivities, such as beta or alpha decays, are negligible for the current experiments using axion haloscopes for the signal enhancement with a high quality factor $Q$ and fine energy resolution of the readout. Furthermore, we note that broadband designs utilizing photon counters can be compromised by the cosmic muons and natural radioactivities, particularly without sufficient signal enhancement or adequate energy resolution.

\section*{Acknowledgements}
We thank the University of Washington Research Award from the Global Innovation Fund to support the collaboration of this work. 



\bibliographystyle{unsrt}
\bibliography{biblio}

@article{Peccei1977June,
  title = {$\mathrm{CP}$ Conservation in the Presence of Pseudoparticles},
  author = {Peccei, R. D. and Quinn, Helen R.},
  journal = {Phys. Rev. Lett.},
  volume = {38},
  issue = {25},
  pages = {1440--1443},
  numpages = {0},
  year = {1977},
  month = {Jun},
  publisher = {American Physical Society},
  doi = {10.1103/PhysRevLett.38.1440},
  url = {https://link.aps.org/doi/10.1103/PhysRevLett.38.1440}
}

@article{Weinberg:1977ma,
      author         = "Weinberg, Steven",
      title          = "{A New Light Boson?}",
      journal        = "Phys. Rev. Lett.",
      volume         = "40",
      pages          = "223-226",
      doi            = "10.1103/PhysRevLett.40.223",
      year           = "1978",
      reportNumber   = "HUTP-77/A074",
      SLACcitation   = "%%CITATION = PRLTA,40,223;%%",
}

@article{Wilczek:1977pj,
      author         = "Wilczek, Frank",
      title          = "{Problem of Strong p and t Invariance in the Presence of
                        Instantons}",
      journal        = "Phys. Rev. Lett.",
      volume         = "40",
      pages          = "279-282",
      doi            = "10.1103/PhysRevLett.40.279",
      year           = "1978",
      reportNumber   = "Print-77-0939 (COLUMBIA)",
      SLACcitation   = "%%CITATION = PRLTA,40,279;%%",
}

@article{ABBOTT1983133,
title = {A cosmological bound on the invisible axion},
journal = {Physics Letters B},
volume = {120},
number = {1},
pages = {133-136},
year = {1983},
issn = {0370-2693},
doi = {https://doi.org/10.1016/0370-2693(83)90638-X},
url = {https://www.sciencedirect.com/science/article/pii/037026938390638X},
author = {L.F. Abbott and P. Sikivie}
}

@article{PhysRevLett.50.925,
  title = {Can Galactic Halos Be Made of Axions?},
  author = {Ipser, J. and Sikivie, P.},
  journal = {Phys. Rev. Lett.},
  volume = {50},
  issue = {12},
  pages = {925--927},
  numpages = {0},
  year = {1983},
  month = {Mar},
  publisher = {American Physical Society},
  doi = {10.1103/PhysRevLett.50.925},
  url = {https://link.aps.org/doi/10.1103/PhysRevLett.50.925}
}

@article{PRESKILL1983127,
title = {Cosmology of the invisible axion},
journal = {Physics Letters B},
volume = {120},
number = {1},
pages = {127-132},
year = {1983},
issn = {0370-2693},
doi = {https://doi.org/10.1016/0370-2693(83)90637-8},
url = {https://www.sciencedirect.com/science/article/pii/0370269383906378},
author = {John Preskill and Mark B. Wise and Frank Wilczek}
}

@article{DINE1983137,
title = {The not-so-harmless axion},
journal = {Physics Letters B},
volume = {120},
number = {1},
pages = {137-141},
year = {1983},
issn = {0370-2693},
doi = {https://doi.org/10.1016/0370-2693(83)90639-1},
url = {https://www.sciencedirect.com/science/article/pii/0370269383906391},
author = {Michael Dine and Willy Fischler}
}

@article{Kim:1979if,
      author         = "Kim, Jihn E.",
      title          = "{Weak Interaction Singlet and Strong CP Invariance}",
      journal        = "Phys. Rev. Lett.",
      volume         = "43",
      pages          = "103",
      doi            = "10.1103/PhysRevLett.43.103",
      year           = "1979",
      reportNumber   = "UPR-0120T",
      SLACcitation   = "%%CITATION = PRLTA,43,103;%%",
}

@article{Shifman:1979if,
      author         = "Shifman, Mikhail A. and Vainshtein, A.I. and Zakharov,
                        Valentin I.",
      title          = "{Can Confinement Ensure Natural CP Invariance of Strong
                        Interactions?}",
      journal        = "Nucl.Phys.",
      volume         = "B166",
      pages          = "493",
      doi            = "10.1016/0550-3213(80)90209-6",
      year           = "1980",
      reportNumber   = "ITEP-64-1979",
      SLACcitation   = "%%CITATION = NUPHA,B166,493;%%",
}

@article{Dine:1981rt,
      author         = "Dine, Michael and Fischler, Willy and Srednicki, Mark",
      title          = "{A Simple Solution to the Strong CP Problem with a        Harmless Axion}",
      journal        = "Phys. Lett.",
      volume         = "B104",
      pages          = "199",
      doi            = "10.1016/0370-2693(81)90590-6",
      year           = "1981",
      reportNumber   = "Print-81-0320 (IAS,PRINCETON)",
      SLACcitation   = "%%CITATION = PHLTA,B104,199;%%",
}

@article{Zhitnitsky:1980tq,
      author         = "Zhitnitsky, A.R.",
      title          = "{On Possible Suppression of the Axion Hadron
                        Interactions. (In Russian)}",
      journal        = "Sov. J. Nucl. Phys.",
      volume         = "31",
      pages          = "260",
      year           = "1980",
      SLACcitation   = "%%CITATION = SJNCA,31,260;%%",
}

@article{PhysRevD.96.095001,
  title = {Axions, instantons, and the lattice},
  author = {Dine, Michael and Draper, Patrick and Stephenson-Haskins, Laurel and Xu, Di},
  journal = {Phys. Rev. D},
  volume = {96},
  issue = {9},
  pages = {095001},
  numpages = {8},
  year = {2017},
  month = {Nov},
  publisher = {American Physical Society},
  doi = {10.1103/PhysRevD.96.095001},
  url = {https://link.aps.org/doi/10.1103/PhysRevD.96.095001}
}

@Article{Borsanyi2016,
author={Borsanyi, S.
and Fodor, Z.
and Guenther, J.
and others},
title={Calculation of the axion mass based on high-temperature lattice quantum chromodynamics},
journal={Nature},
year={2016},
month={Nov},
day={01},
volume={539},
number={7627},
pages={69-71},
issn={1476-4687},
doi={10.1038/nature20115},
url={https://doi.org/10.1038/nature20115}
}

@article{BARDEEN1978229,
title = {Current algebra applied to properties of the light Higgs boson},
journal = {Physics Letters B},
volume = {74},
number = {3},
pages = {229-232},
year = {1978},
issn = {0370-2693},
doi = {https://doi.org/10.1016/0370-2693(78)90560-9},
url = {https://www.sciencedirect.com/science/article/pii/0370269378905609},
author = {William A. Bardeen and S.-H.H. Tye}
}

@article{PhysRevD.18.1607,
  title = {Do axions exist?},
  author = {Donnelly, T. W. and Freedman, S. J. and Lytel, R. S. and others},
  journal = {Phys. Rev. D},
  volume = {18},
  issue = {5},
  pages = {1607--1620},
  numpages = {0},
  year = {1978},
  month = {Sep},
  publisher = {American Physical Society},
  doi = {10.1103/PhysRevD.18.1607},
  url = {https://link.aps.org/doi/10.1103/PhysRevD.18.1607}
}

@article{Dolan_2022,
doi = {10.1088/1475-7516/2022/10/096},
url = {https://doi.org/10.1088/1475-7516/2022/10/096},
year = {2022},
month = {oct},
publisher = {IOP Publishing},
volume = {2022},
number = {10},
pages = {096},
author = {Dolan, Matthew J. and Hiskens, Frederick J. and Volkas, Raymond R.},
title = {Advancing globular cluster constraints on the axion-photon coupling},
journal = {Journal of Cosmology and Astroparticle Physics}
}

@article{PhysRevLett.113.191302,
  title = {Revisiting the Bound on Axion-Photon Coupling from Globular Clusters},
  author = {Ayala, Adrian and Dom\'{\i}nguez, Inma and Giannotti, Maurizio and others},
  journal = {Phys. Rev. Lett.},
  volume = {113},
  issue = {19},
  pages = {191302},
  numpages = {5},
  year = {2014},
  month = {Nov},
  publisher = {American Physical Society},
  doi = {10.1103/PhysRevLett.113.191302},
  url = {https://link.aps.org/doi/10.1103/PhysRevLett.113.191302}
}

@misc{pinetti2025constraintsqcdaxiondark,
      title={First constraints on QCD axion dark matter using James Webb Space Telescope observations}, 
      author={Elena Pinetti},
      year={2025},
      eprint={2503.11753},
      archivePrefix={arXiv},
      primaryClass={hep-ph},
      url={https://arxiv.org/abs/2503.11753}, 
      note={arXiv:2503.11753}
}

@article{PhysRevD.91.065014,
  title = {Axion dark matter from topological defects},
  author = {Kawasaki, Masahiro and Saikawa, Ken'ichi and Sekiguchi, Toyokazu},
  journal = {Phys. Rev. D},
  volume = {91},
  issue = {6},
  pages = {065014},
  numpages = {25},
  year = {2015},
  month = {Mar},
  publisher = {American Physical Society},
  doi = {10.1103/PhysRevD.91.065014},
  url = {https://link.aps.org/doi/10.1103/PhysRevD.91.065014}
}

@article{PhysRevD.83.123531,
  title = {Improved estimation of radiated axions from cosmological axionic strings},
  author = {Hiramatsu, Takashi and Kawasaki, Masahiro and Sekiguchi, Toyokazu and others},
  journal = {Phys. Rev. D},
  volume = {83},
  issue = {12},
  pages = {123531},
  numpages = {12},
  year = {2011},
  month = {Jun},
  publisher = {American Physical Society},
  doi = {10.1103/PhysRevD.83.123531},
  url = {https://link.aps.org/doi/10.1103/PhysRevD.83.123531}
}

@article{PhysRevD.92.034507,
  title = {Lattice QCD input for axion cosmology},
  author = {Berkowitz, Evan and Buchoff, Michael I. and Rinaldi, Enrico},
  journal = {Phys. Rev. D},
  volume = {92},
  issue = {3},
  pages = {034507},
  numpages = {15},
  year = {2015},
  month = {Aug},
  publisher = {American Physical Society},
  doi = {10.1103/PhysRevD.92.034507},
  url = {https://link.aps.org/doi/10.1103/PhysRevD.92.034507}
}

@article{Fleury_2016,
	doi = {10.1088/1475-7516/2016/01/004},
	url = {https://doi.org/10.1088/1475-7516/2016/01/004},
	year = 2016,
	month = {jan},
	publisher = {{IOP} Publishing},
	volume = {2016},
	number = {01},
	pages = {004--004},
	author = {Leesa Fleury and Guy D. Moore},
	title = {Axion dark matter: strings and their cores},
	journal = {Journal of Cosmology and Astroparticle Physics}
}

@Article{Bonati2016,
author={Bonati, Claudio
and D'Elia, Massimo
and Mariti, Marco
and others},
title={Axion phenomenology and $\theta$-dependence from Nf= 2 + 1 lattice QCD},
journal={Journal of High Energy Physics},
year={2016},
month={Mar},
day={22},
volume={2016},
number={3},
pages={155},
issn={1029-8479},
doi={10.1007/JHEP03(2016)155},
url={https://doi.org/10.1007/JHEP03(2016)155}
}

@article{PETRECZKY2016498,
title = {The topological susceptibility in finite temperature QCD and axion cosmology},
journal = {Physics Letters B},
volume = {762},
pages = {498-505},
year = {2016},
issn = {0370-2693},
doi = {https://doi.org/10.1016/j.physletb.2016.09.063},
url = {https://www.sciencedirect.com/science/article/pii/S0370269316305718},
author = {Peter Petreczky and Hans-Peter Schadler and Sayantan Sharma}
}

@article{PhysRevLett.118.071802,
  title = {Unifying Inflation with the Axion, Dark Matter, Baryogenesis, and the Seesaw Mechanism},
  author = {Ballesteros, Guillermo and Redondo, Javier and Ringwald, Andreas and Tamarit, Carlos},
  journal = {Phys. Rev. Lett.},
  volume = {118},
  issue = {7},
  pages = {071802},
  numpages = {7},
  year = {2017},
  month = {Feb},
  publisher = {American Physical Society},
  doi = {10.1103/PhysRevLett.118.071802},
  url = {https://link.aps.org/doi/10.1103/PhysRevLett.118.071802}
}

@article{Klaer_2017,
	doi = {10.1088/1475-7516/2017/11/049},
	url = {https://doi.org/10.1088/1475-7516/2017/11/049},
	year = 2017,
	month = {nov},
	publisher = {{IOP} Publishing},
	volume = {2017},
	number = {11},
	pages = {049--049},
	author = {Vincent B. Klaer and Guy D. Moore},
	title = {The dark-matter axion mass},
	journal = {Journal of Cosmology and Astroparticle Physics}
}

@article{PhysRevLett.124.161103,
  title = {Early-Universe Simulations of the Cosmological Axion},
  author = {Buschmann, Malte and Foster, Joshua W. and Safdi, Benjamin R.},
  journal = {Phys. Rev. Lett.},
  volume = {124},
  issue = {16},
  pages = {161103},
  numpages = {6},
  year = {2020},
  month = {Apr},
  publisher = {American Physical Society},
  doi = {10.1103/PhysRevLett.124.161103},
  url = {https://link.aps.org/doi/10.1103/PhysRevLett.124.161103}
}

@Article{10.21468/SciPostPhys.10.2.050,
	title={{More Axions from Strings}},
	author={Marco Gorghetto and Edward Hardy and Giovanni Villadoro},
	journal={SciPost Phys.},
	volume={10},
	issue={2},
	pages={50},
	year={2021},
	publisher={SciPost},
	doi={10.21468/SciPostPhys.10.2.050},
	url={https://scipost.org/10.21468/SciPostPhys.10.2.050},
}

@article{buschmann2021dark,
    author = "Buschmann, Malte and Foster, Joshua W. and Hook, Anson and others",
    title = "{Dark matter from axion strings with adaptive mesh refinement}",
    doi = "10.1038/s41467-022-28669-y",
    journal = "Nature Commun.",
    volume = "13",
    number = "1",
    pages = "1049",
    year = "2022"
}

@article{PhysRevD.98.035017,
  title = {Stochastic axion scenario},
  author = {Graham, Peter W. and Scherlis, Adam},
  journal = {Phys. Rev. D},
  volume = {98},
  issue = {3},
  pages = {035017},
  numpages = {16},
  year = {2018},
  month = {Aug},
  publisher = {American Physical Society},
  doi = {10.1103/PhysRevD.98.035017},
  url = {https://link.aps.org/doi/10.1103/PhysRevD.98.035017}
}

@article{PhysRevD.98.015042,
  title = {QCD axion window and low-scale inflation},
  author = {Takahashi, Fuminobu and Yin, Wen and Guth, Alan H.},
  journal = {Phys. Rev. D},
  volume = {98},
  issue = {1},
  pages = {015042},
  numpages = {9},
  year = {2018},
  month = {Jul},
  publisher = {American Physical Society},
  doi = {10.1103/PhysRevD.98.015042},
  url = {https://link.aps.org/doi/10.1103/PhysRevD.98.015042}
}

@article{sikivie1983,
  title = {Experimental Tests of the "Invisible" Axion},
  author = {Sikivie, P.},
  journal = {Phys. Rev. Lett.},
  volume = {51},
  issue = {16},
  pages = {1415--1417},
  numpages = {0},
  year = {1983},
  month = {Oct},
  publisher = {American Physical Society},
  doi = {10.1103/PhysRevLett.51.1415},
  url = {https://link.aps.org/doi/10.1103/PhysRevLett.51.1415}
}

@article{admx2018,
  title = {Search for Invisible Axion Dark Matter with the Axion Dark Matter Experiment},
  author = {Du, N. and Force, N. and Khatiwada, R. and others},
  collaboration = {ADMX Collaboration},
  journal = {Phys. Rev. Lett.},
  volume = {120},
  issue = {15},
  pages = {151301},
  numpages = {5},
  year = {2018},
  month = {Apr},
  publisher = {American Physical Society},
  doi = {10.1103/PhysRevLett.120.151301},
  url = {https://link.aps.org/doi/10.1103/PhysRevLett.120.151301}
}

@article{admx2020,
  title = {Extended Search for the Invisible Axion with the Axion Dark Matter Experiment},
  author = {Braine, T. and Cervantes, R. and Crisosto, N.  and others},
  collaboration = {ADMX Collaboration},
  journal = {Phys. Rev. Lett.},
  volume = {124},
  issue = {10},
  pages = {101303},
  numpages = {6},
  year = {2020},
  month = {Mar},
  publisher = {American Physical Society},
  doi = {10.1103/PhysRevLett.124.101303},
  url = {https://link.aps.org/doi/10.1103/PhysRevLett.124.101303}
}

@article{admx2021,
  title = {Search for Invisible Axion Dark Matter in the $3.3--4.2\text{ }\text{ }\ensuremath{\mu}\mathrm{eV}$ Mass Range},
  author = {Bartram, C. and Braine, T. and Burns, E. and others},
  collaboration = {ADMX Collaboration},
  journal = {Phys. Rev. Lett.},
  volume = {127},
  issue = {26},
  pages = {261803},
  numpages = {6},
  year = {2021},
  month = {Dec},
  publisher = {American Physical Society},
  doi = {10.1103/PhysRevLett.127.261803},
  url = {https://link.aps.org/doi/10.1103/PhysRevLett.127.261803}
}

@article{admx2024,
  title = {ADMX Axion Dark Matter Bounds around $3.3\text{ }\text{ }\mathrm{\ensuremath{\mu}}\mathrm{eV}$ with Dine-Fischler-Srednicki-Zhitnitsky Discovery Ability},
  author = {Goodman, C. and Guzzetti, M. and Hanretty, C. and others},
  collaboration = {ADMX Collaboration},
  journal = {Phys. Rev. Lett.},
  volume = {134},
  issue = {11},
  pages = {111002},
  numpages = {7},
  year = {2025},
  month = {Mar},
  publisher = {American Physical Society},
  doi = {10.1103/PhysRevLett.134.111002},
  url = {https://link.aps.org/doi/10.1103/PhysRevLett.134.111002}
}

@article{admx2025,
  title = {Search for Axion Dark Matter from 1.1 to 1.3 GHz with ADMX},
  author = {Carosi, G. and Cisneros, C. and Du, N. and others},
  collaboration = {ADMX Collaboration},
  journal = {Phys. Rev. Lett.},
  volume = {135},
  issue = {19},
  pages = {191001},
  numpages = {9},
  year = {2025},
  month = {Nov},
  publisher = {American Physical Society},
  doi = {10.1103/d7mg-6sqq},
  url = {https://link.aps.org/doi/10.1103/d7mg-6sqq}
}

@article{capp2023,
  title = {Axion Dark Matter Search around $4.55\text{ }\text{ }\mathrm{\ensuremath{\mu}}\mathrm{eV}$ with Dine-Fischler-Srednicki-Zhitnitskii Sensitivity},
  author = {Yi, Andrew K. and Ahn, Saebyeok and Kutlu, \ifmmode \mbox{\c{C}}\else \c{C}\fi{}a\ifmmode \breve{g}\else \u{g}\fi{}lar and others},
  journal = {Phys. Rev. Lett.},
  volume = {130},
  issue = {7},
  pages = {071002},
  numpages = {7},
  year = {2023},
  month = {Feb},
  publisher = {American Physical Society},
  doi = {10.1103/PhysRevLett.130.071002},
  url = {https://link.aps.org/doi/10.1103/PhysRevLett.130.071002}
}

@article{capp2024,
  title = {Extensive Search for Axion Dark Matter over 1 GHz with CAPP'S Main Axion Experiment},
  author = {Ahn, Saebyeok and Kim, JinMyeong and Ivanov, Boris I. and others},
  journal = {Phys. Rev. X},
  volume = {14},
  issue = {3},
  pages = {031023},
  numpages = {32},
  year = {2024},
  month = {Aug},
  publisher = {American Physical Society},
  doi = {10.1103/PhysRevX.14.031023},
  url = {https://link.aps.org/doi/10.1103/PhysRevX.14.031023}
}

@article{PhysRevD.111.092012,
  title = {Improved receiver noise calibration for ADMX axion search: 4.54 to $5.41\text{ }\text{ }\mathrm{\ensuremath{\mu}eV}$},
  author = {Guzzetti, M. and Zhang, D. and Goodman, C. and others},
  collaboration = {ADMX Collaboration},
  journal = {Phys. Rev. D},
  volume = {111},
  issue = {9},
  pages = {092012},
  numpages = {13},
  year = {2025},
  month = {May},
  publisher = {American Physical Society},
  doi = {10.1103/PhysRevD.111.092012},
  url = {https://link.aps.org/doi/10.1103/PhysRevD.111.092012}
}

@article{geant4_1,
title = {Recent developments in Geant4},
journal = {Nuclear Instruments and Methods in Physics Research Section A: Accelerators, Spectrometers, Detectors and Associated Equipment},
volume = {835},
pages = {186-225},
year = {2016},
issn = {0168-9002},
doi = {https://doi.org/10.1016/j.nima.2016.06.125},
url = {https://www.sciencedirect.com/science/article/pii/S0168900216306957},
author = {J. Allison and K. Amako and J. Apostolakis and others},
}

@ARTICLE{geant4_2,
  author={Allison, J. and Amako, K. and Apostolakis, J. and others},
  journal={IEEE Transactions on Nuclear Science}, 
  title={Geant4 developments and applications}, 
  year={2006},
  volume={53},
  number={1},
  pages={270-278},
  doi={10.1109/TNS.2006.869826}}

@article{geant4_3,
title = {Geant4—a simulation toolkit},
journal = {Nuclear Instruments and Methods in Physics Research Section A: Accelerators, Spectrometers, Detectors and Associated Equipment},
volume = {506},
number = {3},
pages = {250-303},
year = {2003},
issn = {0168-9002},
doi = {https://doi.org/10.1016/S0168-9002(03)01368-8},
url = {https://www.sciencedirect.com/science/article/pii/S0168900203013688},
author = {S. Agostinelli and J. Allison and K. Amako and J. Apostolakis and others},
}

@book{1966,
    author = {G. BEKEFI},
    title = {Radiation Processes In Plasmas},
    publisher = {JOHN WILEY AND SONS},
    year = {1966}
}

@article{cosmicMuonSource,
title = {EcoMug: An Efficient COsmic MUon Generator for cosmic-ray muon applications},
journal = {Nuclear Instruments and Methods in Physics Research Section A: Accelerators, Spectrometers, Detectors and Associated Equipment},
volume = {1014},
pages = {165732},
year = {2021},
issn = {0168-9002},
doi = {https://doi.org/10.1016/j.nima.2021.165732},
url = {https://www.sciencedirect.com/science/article/pii/S0168900221007178},
author = {D. Pagano and G. Bonomi and A. Donzella and A. Zenoni and others}
}

@article{admxDetails,
    author = {Khatiwada, R. and Bowring, D. and Chou, A. S. and others},
    title = {Axion Dark Matter Experiment: Detailed design and operations},
    journal = {Review of Scientific Instruments},
    volume = {92},
    number = {12},
    pages = {124502},
    year = {2021},
    month = {12},
    issn = {0034-6748},
    doi = {10.1063/5.0037857},
    url = {https://doi.org/10.1063/5.0037857},
    eprint = {https://pubs.aip.org/aip/rsi/article-pdf/doi/10.1063/5.0037857/16136265/124502_1_online.pdf}
}

@book{lienard,
  author = {Jackson, J. D.},
  title = {Classical Electrodynamics},
  edition = {3rd},
  publisher = {Wiley},
  year = {1999},
  pages = {664-668}

}

@article{darkMatterDensity_1,
doi = {10.1086/309652},
url = {https://doi.org/10.1086/309652},
year = {1995},
month = {aug},
publisher = {},
volume = {449},
number = {2},
pages = {L123},
author = {Gates, Evalyn I. and Gyuk, Geza and Turner, Michael S.},
title = {The Local Halo Density},
journal = {The Astrophysical Journal}
}

@article{darkMatterDensity_2,
doi = {10.1088/1361-6633/ac24e7},
url = {https://doi.org/10.1088/1361-6633/ac24e7},
year = {2021},
month = {oct},
publisher = {IOP Publishing},
volume = {84},
number = {10},
pages = {104901},
author = {de Salas, Pablo F and Widmark, A},
title = {Dark matter local density determination: recent observations and future prospects},
journal = {Reports on Progress in Physics}
}

@article{madmax,
  title = {First Search for Axion Dark Matter with a MADMAX Prototype},
  author = {Ary dos Santos Garcia, B. and Bergermann, D. and Caldwell, A. and others},
  collaboration = {MADMAX Collaboration},
  journal = {Phys. Rev. Lett.},
  volume = {135},
  issue = {4},
  pages = {041001},
  numpages = {7},
  year = {2025},
  month = {Jul},
  publisher = {American Physical Society},
  doi = {10.1103/c749-419q},
  url = {https://link.aps.org/doi/10.1103/c749-419q}
}

@article{bread,
  title = {First Axionlike Particle Results from a Broadband Search for Wavelike Dark Matter in the 44 to $52\text{ }\text{ }\mathrm{\ensuremath{\mu}}\mathrm{eV}$ Range with a Coaxial Dish Antenna},
  author = {Hoshino, Gabe and Knirck, Stefan and Awida, Mohamed H. and others},
  collaboration = {GigaBREAD Collaboration},
  journal = {Phys. Rev. Lett.},
  volume = {134},
  issue = {17},
  pages = {171002},
  numpages = {9},
  year = {2025},
  month = {Apr},
  publisher = {American Physical Society},
  doi = {10.1103/PhysRevLett.134.171002},
  url = {https://link.aps.org/doi/10.1103/PhysRevLett.134.171002}
}

@misc{gavnet,
      title={The Global Network of Cavities to Search for Gravitational Waves (GravNet): A novel scheme to hunt gravitational waves signatures from the early universe}, 
      author={Kristof Schmieden and Matthias Schott},
      year={2023},
      eprint={2308.11497},
      archivePrefix={arXiv},
      primaryClass={gr-qc},
      url={https://arxiv.org/abs/2308.11497}, 
      note={arXiv:2308.11497}
}

@article{project8,
  title = {Antenna arrays for neutrino mass measurements with cyclotron radiation emission spectroscopy},
  author = {Ashtari Esfahani, A. and Bhagvati, S. and B\"oser, S. and Brandsema, M. J. and others},
  collaboration = {Project 8 Collaboration},
  journal = {Phys. Rev. C},
  volume = {112},
  issue = {4},
  pages = {045506},
  numpages = {25},
  year = {2025},
  month = {Oct},
  publisher = {American Physical Society},
  doi = {10.1103/dc29-s4y2},
  url = {https://link.aps.org/doi/10.1103/dc29-s4y2}
}

@article{helium6,
    author = "Buzinsky, N. and others",
    title = "{Larmor power limit for cyclotron radiation of relativistic particles in a waveguide}",
    eprint = "2405.06847",
    archivePrefix = "arXiv",
    primaryClass = "nucl-ex",
    doi = "10.1088/1367-2630/ad6d85",
    journal = "New J. Phys.",
    volume = "26",
    number = "8",
    pages = "083021",
    year = "2024"
}

\end{document}